\documentclass[12pt]{article}
\usepackage[margin=1in]{geometry}
\usepackage{graphicx} 
\usepackage{amsmath}
\usepackage{amssymb}
\usepackage{amsthm}
\usepackage{rotating}
\usepackage{amsfonts}
\usepackage[labelfont=bf]{caption}
\usepackage[usenames]{color}
\usepackage{bm}
\usepackage[breaklinks=true,colorlinks,citecolor=black,linkcolor=blue,urlcolor=black]{hyperref}
\usepackage{breakcites}
\usepackage[authoryear,round]{natbib}
\usepackage[capitalize,noabbrev]{cleveref}
\usepackage{enumitem}
\usepackage{dsfont}
\usepackage{makecell}
\graphicspath{ {figures/} }

\theoremstyle{plain}
\newtheorem{theorem}{Theorem}[section]
\newtheorem{algorithm}[theorem]{Algorithm}

\DeclareMathOperator*{\argmin}{argmin}

\title{Principal variables analysis for non-Gaussian data}
\author{
  Dylan Clark-Boucher\\
  \texttt{dclarkboucher@fas.harvard.edu}
  \and
  Jeffrey W. Miller\\
  \texttt{jwmiller@hsph.harvard.edu}
}

\date{}

\begin{document}

\maketitle 

\begin{abstract}
Principal variables analysis (PVA) is a technique for selecting a subset of variables that capture as much of the information in a dataset as possible.
Existing approaches for PVA are based on the Pearson correlation matrix, which is not well-suited to describing the relationships between non-Gaussian variables. We propose a generalized approach to PVA enabling the use of different types of correlation,
and we explore using Spearman, Gaussian copula, and polychoric correlations as alternatives to Pearson correlation when performing PVA.
We compare performance in simulation studies varying the form of the true multivariate distribution over a wide range of possibilities.
Our results show that on continuous non-Gaussian data, using generalized PVA with Gaussian copula or Spearman correlations provides a major improvement in performance compared to Pearson.
Meanwhile, on ordinal data, generalized PVA with polychoric correlations outperforms the rest by a wide margin.
We apply generalized PVA to a dataset of 102 clinical variables measured on individuals with X-linked dystonia parkinsonism (XDP), a rare neurodegenerative disorder, and we find that using different types of correlation yields substantively different sets of principal variables.

\end{abstract}

\section{Introduction}
Principal variables analysis (PVA) is technique for selecting a subset of variables that retain the properties of the complete data as well as possible. Different variants of PVA have be used to reduce dimensionality in a way that retains the structure of the original feature space, preserves the relative distance between data points, or explains the variability of features that are not chosen. 
While principal variables analysis is closely related to principal components analysis (PCA), the main difference is that PVA selects a subset of variables, rather than linear combinations of variables \citep{McCabe,Beale}. In comparison to PCA, PVA is particularly useful for deciding which variables are worth collecting in future studies.

Approaches to PVA have tended to take on three forms. Criterion-based approaches, such as those presented by \citet{McCabe} and \citet{Cadima}, apply best subset algorithms or sequential greedy algorithms to maximize a pre-defined optimality criterion \citep{Brusco, Cumming}. PCA-based approaches, such as those discussed by \citet{Jolliffe72,Jolliffe73}, select variables sequentially based on their contributions to important principal components. Clustering-based approaches identify clusters of correlated variables, then select one or more variables from each cluster so that each cluster is represented in the reduced data. 
Despite their differences, all of these methods share a key feature, which is that they are based on the sample Pearson correlation matrix. For instance, an appealing criterion-based approach is to minimize the trace of the conditional covariance matrix of the omitted variables given the chosen variables, while holding the number of chosen variables constant \citep{McCabe}. If the variables are multivariate Gaussian, that is, multivariate normal (MVN), and have been normalized to unit variance, this is equivalent to maximizing the ``variance explained" --- that is, the amount of variance of the omitted variables explained by the chosen variables.

However, a major limitation of these existing methods is that on non-Gaussian data, the Pearson correlation is not a very natural measure of interdependence. For example, it is well-known that two binary variables cannot have correlation 1 unless their marginal probabilities are equal. 
Similarly, variables may be highly dependent in an unobserved latent space, but exhibit lower correlation in the observed data due to discretization or other transformations that occur in the measurement process. This can cause PVA methods based on the Pearson correlation to miss out on important features.

In this article, we propose a generalized PVA method for handling non-Gaussian data, including ordinal and continuous cases. 
The basic idea is simply to replace the Pearson correlation matrix with an alternative correlation matrix when running PVA.
We investigate using three alternatives: (1) Spearman correlation, which is based on the observed ranks; (2) polychoric correlation, which is designed for ordinal variables \citep{Choi}; and (3) Gaussian copula correlation, which can capture arbitrary latent correlation structure and arbitrary marginals \citep{Trivedi}. We evaluate each  approach under diverse simulation conditions with varying assumptions about the true multivariate distribution. We focus on the criterion-based method for PVA described above, in which we seek to minimize the trace of the conditional covariance matrix of the omitted variables given the chosen variables. To the best of our knowledge, this is the first study to directly examine the limitations of using the Pearson correlation matrix for PVA, while presenting alternatives that are robust to changes in the true data-generating mechanism.

The article is organized as follows. In \cref{sec:methods}, we describe the PVA algorithm based on the residual trace criterion, and we extend the method beyond Pearson to Spearman, polychoric, and Gaussian copula correlations. In \cref{sec:simulations}, we evaluate the performance of these methods on Gaussian and non-Gaussian data in a range of simulation studies. In \cref{sec:application}, we apply the methods to a dataset containing 102 clinical variables from individuals with X-linked dystonia parkinsonism (XDP), a rare and under-studied genetic disease. We conclude in \cref{sec:discussion} with a brief discussion.

\section{Methods}
\label{sec:methods}

\subsection{PVA algorithm}

The variant of PVA we will consider is based on the McCabe ``variance-explained" criterion, which selects the subset of variables that explains as much of the variance in the other variables as possible. Let $X$ denote a random vector of length $p$ and let $S \subseteq \{1,\ldots,p\}$ such that $|S| = q$. We write $X_S$ to denote the sub-vector $X_S = (X_j : j \in S) \in \mathbb{R}^q$, and $X_{S^c}$ to denote the complementary sub-vector $X_{S^c} = (X_j : j \not\in S) \in \mathbb{R}^{p-q}$. Then the optimal subset is defined as
\begin{align}
\label{eq:mccabe-criterion}
S^* = \argmin_{S\, :\, |S|=q}\mathrm{tr}\big(\mathrm{Cov}(X_{S^c}\mid X_S)\big)
\end{align}
where $\mathrm{tr}(\cdot)$ is the trace of a matrix. Other criteria for PVA have considered minimizing the determinant of $\mathrm{Cov}(X_{S^c}\mid X_S)$ or the Frobenius norm \citep{McCabe,Cumming}. 

The covariance matrix in \cref{eq:mccabe-criterion} is straightforward to compute when the data follow a multivariate Gaussian distribution. If $[X_S^\texttt{T},X_{S^c}^\texttt{T}]^\texttt{T}\sim\mathrm{MVN}(\bm{\mu}$, $\bm{{\Sigma})}$ where
\begin{gather*}
    \bm{\Sigma} = 
    \begin{bmatrix}
    \bm{\Sigma}_{11} & \bm{\Sigma}_{12} \\
    \bm{\Sigma}_{21} & \bm{\Sigma}_{22}
    \end{bmatrix}
\end{gather*}
such that $\Sigma_{11} = \mathrm{Cov}(X_S)$ and $\Sigma_{22} = \mathrm{Cov}(X_{S^c})$, then the conditional covariance matrix of $X_{S^c}$ given $X_S$ is
\begin{align}
\label{eq:conditional-covariance-full}
    \bm\Sigma_{S^c\mid S} := \mathrm{Cov}(X_{S^c}\mid X_S) = \bm{\Sigma}_{22} - \bm{\Sigma}_{21}\bm{\Sigma}_{11}^{-1}\bm{\Sigma}_{12}.
\end{align}
As a special case of this formula, if $X_j$ is the $j$th entry of $X$, and $-j$ denotes $\{j\}^c$, then the conditional covariance of all the other variables in $X$ given $X_j$ is given by
\begin{align}
\label{eq:conditional-covariance-single}
    \bm\Sigma_{-j\mid j} := \mathrm{Cov}(X_{-j} \mid X_j) = \bm{\Sigma}_{-j} - \bm{\Sigma}_{-j,j}\bm{\Sigma}_{-j,j}^\texttt{T} / \sigma^2_j,
\end{align}
where $\sigma^2_j = \mathrm{Var}(X_j)$, $\bm{\Sigma}_{-j} = \mathrm{Cov}(X_{-j})$, and $\bm{\Sigma}_{-j,j}$ is a vector of the covariance between each entry of $X_{-j}$ and $X_j$.

To avoid a computationally intensive search over sub-vectors of $X$, we employ a greedy algorithm for finding an approximate solution to the optimization problem; see \cref{alg:pva}. Briefly, starting with an estimated covariance matrix, the first step in the algorithm is to find the variable $X_j$ that yields a conditional covariance matrix with the smallest trace based on \cref{eq:conditional-covariance-single}. This $j$ becomes the first index $j_1$ to be included in our selected set $\tilde{S}^* = \{j_1,\ldots,j_q\}$, and the algorithm repeats with $\bm\Sigma_{-j\mid j}$  as the new covariance matrix. 

\begin{algorithm}~
\label{alg:pva}
\begin{enumerate}[itemsep=0pt]
    \item[] Input: $\bm{\Sigma} \in \mathbb{R}^{p\times p}$ positive definite.
    \item[] Output: Indices of selected variables, $j_1,\ldots,j_q$.
    \item[] (1) Initialize $S \leftarrow \{1,\ldots,p\}$.
    \item[] (2) For $k = 1,\ldots,q$:
    \begin{enumerate}
        \item $i \leftarrow \argmin_j \mathrm{tr}(\bm{\Sigma}_{-j\mid j})$
        \item $j_k \leftarrow S_i$
        \item $\bm{\Sigma} \leftarrow \bm{\Sigma}_{-i\mid i}$
        \item $S \leftarrow S \setminus \{j_k\}$.
    \end{enumerate}
\end{enumerate}
\end{algorithm}

Note that this algorithm can be run without directly observing the data, since the only required input is a matrix $\bm{\Sigma}$. In addition, since the scale of each variable is not relevant to the dependency between variables, a natural input to this algorithm is the correlation matrix rather than the covariance matrix. In the next section, we discuss the limitations of using the Pearson correlation matrix for this purpose and consider more robust alternatives for the choice of $\bm{\Sigma}$.

\subsection{Generalized PVA using alternative correlations}

It is widely recognized that Pearson correlation has drawbacks when measuring the interdependence of non-Gaussian data. For example, it is not invariant to monotone transformations of the data, it is constrained by the marginal distributions, and it is incapable of describing bivariate tail dependence. We consider three alternatives to Pearson correlation in the context of PVA: Spearman correlation, polychoric correlation, and Gaussian copula correlation.

Spearman correlation is equivalent to the Pearson correlation of the ranks. Ties are handled by averaging the ranks of each set of tied points and assigning this average rank to those points, which is equivalent to averaging over all permutations of the ties \citep{Dodge}. An attractive feature of Spearman correlation is that it is invariant to monotonically increasing transformations of the data; thus, it is nonparametric with respect to the marginal distributions. 
However, since the ranks of the data are uniformly distributed, the conditional covariance is not given by the formula in \cref{eq:conditional-covariance-full}, since this formula is based on the multivariate Gaussian assumption. 

Polychoric correlation extends the Pearson correlation to ordinal variables by assuming the variables have a latent multivariate Gaussian structure \citep{Choi}. Specifically, suppose $[Z_1, Z_2]^\texttt{T}\in\mathbb{R}^2$ is bivariate Gaussian with unit variances and correlation $\rho$, and $Y_1,Y_2$ are ordinal variables defined by unknown monotonic transformations of $Z_1,Z_2$, respectively. Then $\rho$ is referred to as the polychoric correlation of $Y_1,Y_2$, and it can be estimated via an iterative maximum-likelihood approach given observations of $Y_1,Y_2$ \citep{Olsson_1979}. The estimation procedure can also be applied when we have observations of $Z_1,Y_2$ rather than $Y_1,Y_2$.

Gaussian copulas are tractable models of multivariate dependence with arbitrary marginal distributions, popular in economics \citep{Trivedi}. Suppose the data comprise $n$ observations of a multivariate random vector $X=({X_1,X_2,\dots,X_p})^\textit{T}$, with an unknown cumulative distribution function  $F_X(x_1,x_2,\dots,x_p)$. A Gaussian copula provides an approximation to $F_X$ of the form
\begin{align*}
    \hat{F}_{X}(x_1,x_2,\dots,x_p) = \Phi_p\big(\Phi^{-1}(\hat{F_1}(x_1)),\Phi^{-1}(\hat{F}_2(x_2)),\dots,\Phi^{-1}(\hat{F}_p(x_p))\;\big\vert\; \bm{0},\bm{\hat{\Sigma}}\big),
\end{align*}
where $\Phi$ is the cumulative distribution function (CDF) of a standard normal distribution, $\Phi_p(\cdot\mid \bm{\mu},\bm{\Sigma})$ is the CDF of a multivariate Gaussian distribution with mean $\bm{\mu}$ and covariance matrix  $\bm{\Sigma}$, and $\hat{F}_1,\hat{F}_2,\dots,\hat{F}_p$ are estimated CDFs of the $p$ variables. To fit this model, the approximations $\hat{F}_j$ can be obtained parametrically by specifying a functional form of the distributions and applying maximum likelihood, or non-parametrically by taking the empirical CDF of the observed data. Finally, the estimate $\bm{\hat{\Sigma}}$ can be defined as the sample Pearson correlation matrix of the transformed variables $\Phi^{-1}(\hat{F_1}(X_1)),\Phi^{-1}(\hat{F}_2(X_2)),\dots,\Phi^{-1}(\hat{F}_p(X_p))$. The interpretation of $\bm{\hat{\Sigma}}$ is the covariance matrix of a multivariate Gaussian distribution that can be constructed by applying monotonic transformations to the observed variables. However, the validity of this interpretation depends on the strong assumption that the transformed variables $\Phi^{-1}({F_1}(X_1)),\Phi^{-1}({F}_2(X_2)),\dots,\Phi^{-1}({F}_p(X_p))$ are multivariate Gaussian.

Each of these three approaches (Spearman, polychoric, and Gaussian copula) produces an estimated correlation matrix $\bm{\Sigma}$ that can be used to perform PVA using \cref{alg:pva}. Importantly, the conditional covariance formula in \cref{eq:conditional-covariance-full} is valid for the polychoric and Gaussian copula approaches, since $\bm{\Sigma}$ is the correlation matrix of multivariate Gaussian latent variables in these approaches.  The implied assumption of these  approaches is that we are interested in the dependence structure of the latent variables rather than the observed variables, which may or may not be appropriate depending on the application. In the application in \cref{sec:application}, for instance, this is a natural assumption since the ordinal variables represent continuous traits that were measured on ordinal scales for practical reasons.

\section{Simulations}
\label{sec:simulations}

In this section, we evaluate the performance of PVA using the Spearman, Gaussian copula, and polychoric correlation approaches compared to the standard approach using the Pearson correlation.  To simulate data, we generate latent variables from a multivariate Gaussian distribution and transform them by applying monotonic functions to each variable separately, in order to evaluate the effect of non-Gaussianity on PVA performance. The latent multivariate Gaussian distribution serves as ground truth for defining the ideal value of the optimality criterion that would be used if the latent distribution were known.
We consider a variety of monotonic transformations (yielding a variety of non-Gaussian marginal distributions) and we assess performance in two ways: (1) the proportion of ideal variables selected, and (2) the variance explained by the selected variables, quantified in terms of relative explanatory efficiency.

\subsection{Proportion of ideal variables selected}
\label{sec:simulations-proportion-selected}

For the first set of simulations, we consider examples in which non-Gaussianity causes Pearson correlation-based PVA to miss variables that are important for capturing latent multivariate structure. The general setup is to simulate a latent data matrix $\bm{X} = [X_{i j}]\in\mathbb{R}^{n\times p}$ comprising $n$ points sampled i.i.d.\ from a $\mathrm{MVN}_p(\bm{0}, \bm{\Sigma}$) distribution, and determine an ``ideal" set of $q$ latently important variables by performing PVA on $\bm{\Sigma}$. We then generate 
$\bm{Y}$, the observed data matrix, by the transformation
\begin{gather*}
    Y_{i j} = 
    \begin{cases}
        f_j(X_{i j}) & \text{if } j \in H\\
        X_{i j} & \text{if } j\notin H
    \end{cases}
\end{gather*}
where $f_1,\ldots,f_p$ are monotonic functions and $H\subseteq \{1,2,\dots,p\}$ is a fixed subset of $K$ variables that will be transformed. We evaluate the performance of PVA with different correlation matrices (Pearson, Spearman, Gaussian copula, and polychoric) estimated from the data $\bm{Y}$ by contrasting the sets of variables obtained from those methods with the ideal set of variables obtained by PVA on $\bm{\Sigma}$. The more these two sets of variables overlap, the better PVA is performing.

We run simulations using the following settings.  With $p=10$, we randomly generate $\bm{\Sigma}$ by sampling from a Wishart distribution with $p$ degrees of freedom and scale matrix $I$, then normalize this matrix to have ones on the diagonal so it is a correlation matrix. We then perform PVA on this ground truth $\bm{\Sigma}$ to identify the ideal set of $q=5$ variables, say $j^*_1,\ldots,j^*_5$, that capture the most variance in the latent multivariate distribution.
We then sample latent data $\bm{X}$ from $\mathrm{MVN}_p(\bm{0}, \bm{\Sigma}$), and generate the observed data $\bm{Y}$ by transforming variables $H=\{j^*_1,\ldots,j^*_5\}$ in one of three different ways: (A) no transformation, (B) continuous transformation, or (C) ordinal transformation.  (That is, the five variables that are most important latently are the variables that are transformed). In each case, for each $n\in\{50,150,400,1200,3500,10000\}$, we perform 1000 replicate simulations of the process of generating $\bm{\Sigma}, \bm{X}, \bm{Y}$, running the PVA methods, and comparing the chosen sets of variables to the latently ideal set, $\{j^*_1,\ldots,j^*_5\}$.

(A) \textit{No transformation.} To compare performance when the data are indeed multivariate Gaussian, we run the suite of simulations without any transformation of $\bm{X}$, so that $\bm{Y} = \bm{X}$.  We run PVA using Pearson, Spearman, and Gaussian copula correlations; polychoric is not considered here since it is designed for ordinal variables.

(B) \textit{Continuous transformation.}
In this case, we transform the ideal variables  $j^*_1,\ldots,j^*_5$ with the five monotonically increasing functions
\begin{align*}
Y_{ij^*_1} &= \mathrm{min}(\hat{F}_{j^*_1}(X_{ij^*_1})^2, 0.6^2) \\
Y_{ij^*_2} &= \mathrm{min}(\hat{F}_{j^*_2}(X_{ij^*_2})^6, 0.6^6) \\
Y_{ij^*_3} &=  F^{-1}_{\mathrm{Gamma}}(\hat{F}_{j^*_3}(X_{ij^*_3}),0.5,1)   \\
Y_{ij^*_4} &= F^{-1}_{\mathrm{Pareto}}(\hat{F}_{j^*_4}(X_{ij^*_4}),2,1) \\  
Y_{ij^*_5} &= \exp\big(\hat{F}_{j^*_5}(X_{ij^*_5})\big)[1 + \mathds{1}(\hat{F}_{j^*_5}(X_{ij^*_5}) > 0.9)]
\end{align*}
for $i\in \{1,2,\dots,n\}$, where $\hat{F}_{k}$ is the empirical CDF of $X_{1 k},\ldots,X_{n k}$, times $n/(n+1)$, so that
\begin{align*}
\hat{F}_{k}(X_{ik}) = \frac{\mathrm{Rank}(X_{ik})}{n+1}.
\end{align*}
Here, $\mathrm{Rank}(X_{ik})$ is the ranking (from least to greatest) of $X_{ik}$ in $\bm{X}_{k}$, the $k$th column of $\bm{X}$. (Ties in the rank are handled by taking the average rank over all permutations of the column.) Meanwhile, $\mathds{1}(\cdot)$ is the indicator function, $F^{-1}_{\mathrm{Gamma}}(\cdot,a,b)$ is the inverse CDF of the gamma distribution with shape $a$ and scale $b$, and $F^{-1}_{\mathrm{Pareto}}(\cdot,a,b)$ is the inverse CDF of a Pareto distribution with shape $a$ and scale $b$.  The empirical CDFs transform the variables to be approximately $\mathrm{Uniform}(0,1)$, marginally, after which they are mapped through an inverse CDF to produce one of five marginal distributions. These inverse CDFs are chosen to reduce the Pearson correlation between the ideal variables and the non-ideal variables, while creating marginal distributions that could realistically be observed in practice. For these simulations, we again use Pearson, Spearman, and Gaussian copula correlations, but not polychoric since it is designed for ordinal variables.

(C) \textit{Ordinal transformation.}
In this case, we transform the ideal variables  $j^*_1,\ldots,j^*_5$ from Gaussian to ordinal variables by applying the following functions, respectively:
\begin{align*}
Y_{ij^*_1} &= 1 + \mathds{1}(\hat{F}_{j^*_1}(X_{ij^*_1}) > 0.2) \\
Y_{ij^*_2} &= 1 + \mathds{1}(\hat{F}_{j^*_2}(X_{ij^*_2}) > 0.4) + \mathds{1}(\hat{F}_{j^*_2}(X_{ij^*_2}) > 0.6) \\
Y_{ij^*_3} &= 1 + \mathds{1}(\hat{F}_{j^*_3}(X_{ij^*_3}) > 0.2) + \mathds{1}(\hat{F}_{j^*_3}(X_{ij^*_3}) > 0.3) \\
Y_{ij^*_4} &= 1 + \mathds{1}(\hat{F}_{j^*_4}(X_{ij^*_4}) > 0.3) + \mathds{1}(\hat{F}_{j^*_4}(X_{ij^*_4}) > 0.5)  + \mathds{1}(\hat{F}_{j^*_4}(X_{ij^*_4}) > 0.7)\\
Y_{ij^*_5} &= 1 + \mathds{1}(\hat{F}_{j^*_5}(X_{ij^*_5}) > 0.1) + \mathds{1}(\hat{F}_{j^*_5}(X_{ij^*_5}) > 0.2)  + \mathds{1}(\hat{F}_{j^*_6}(X_{ij^*_5}) > 0.3).
\end{align*}

This makes $Y_{i j_1^*} \in \{1,2\}$, $Y_{i j_2^*}, Y_{i j_3^*}\in \{1,2,3\}$, and $Y_{i j_4^*}, Y_{i j_5^*}\in \{1,2,3,4,5\}$.
Here, we run PVA with polychoric correlations in addition to Pearson, Spearman, and Gaussian copula, since polychoric correlations are designed specifically for the case of ordinal variables.

\begin{figure}[htbp]
\includegraphics[width=\textwidth]{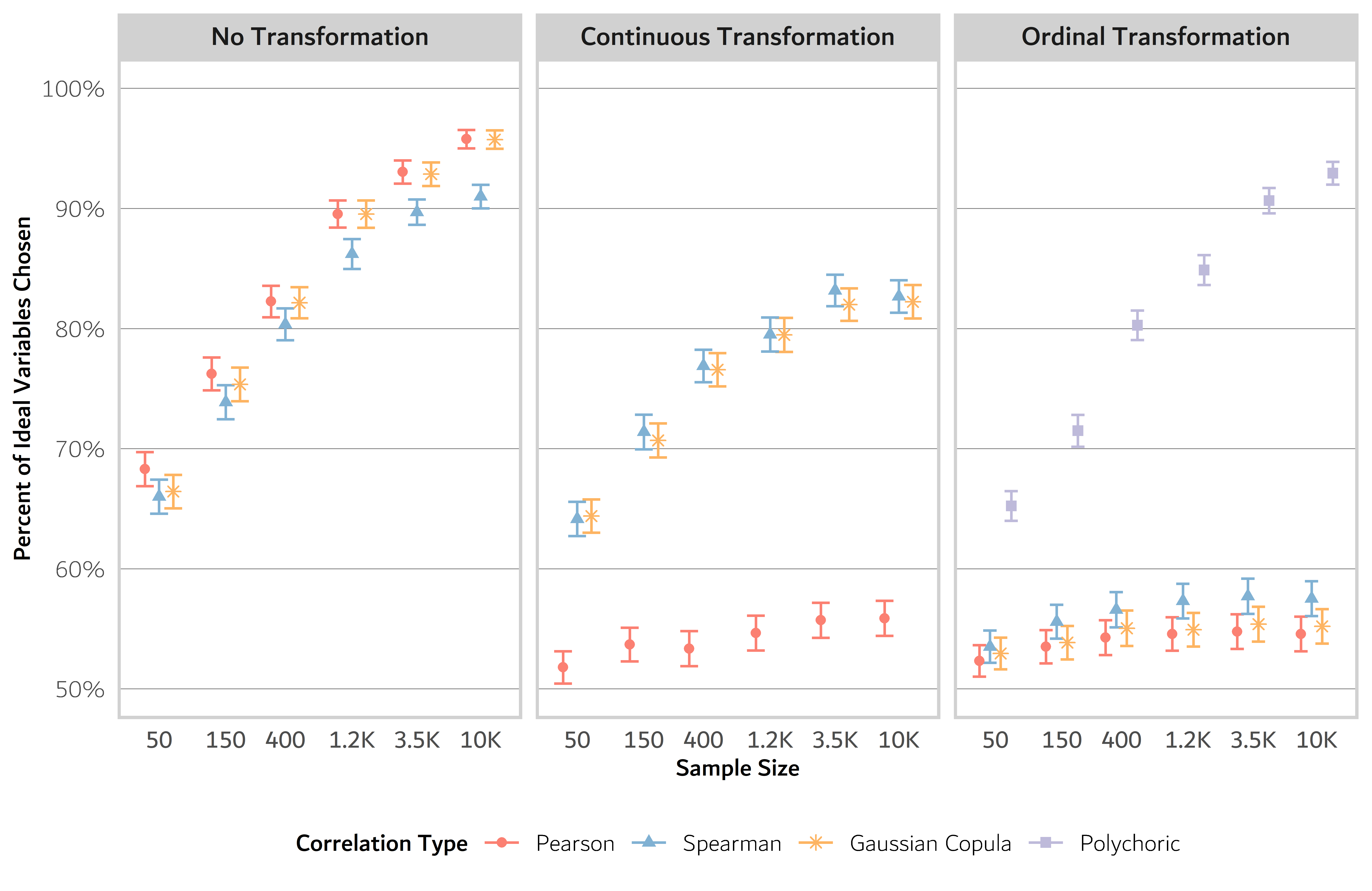}
\caption{\textbf{Proportion of latently ideal variables identified by PVA.} Value shown is the mean proportion over 1,000 replicate simulations, and error bars show $\pm$ two standard errors. The x-axis is on a logarithmic scale.}
\label{fig:simulation-proportion-selected}
\end{figure}

\subsubsection{Results for proportion of ideal variables selected}

Results for these simulations are shown in \cref{fig:simulation-proportion-selected}. The y-axis shows the mean proportion of the latently ideal variables that were selected by PVA when using each method (Pearson, Spearman, Gaussian copula, or polychoric).  In case A (no transformation), where the observed data are Gaussian, there is effectively no difference between the performance of the Pearson and Gaussian copula correlation approaches; both perform very well with the proportion nearing 100\% as the sample size exceeds $10{,}000$. Spearman correlations also perform well but slightly worse. This makes sense since Pearson and Gaussian copula correlations either implicitly or explicitly employ Gaussian assumptions, whereas Spearman is a fully nonparametric approach that is expected to entail some loss of information.

In case B (continuous transformation), Spearman and Gaussian copula correlations perform moderately well---with the proportion of ideal variables chosen exceeding 80\%---whereas Pearson is unable to reach even 60\%. This accords with intuition since the Spearman and Gaussian copula approaches are designed to handle non-Gaussian marginals, whereas Pearson is not robust to departures from Gaussianity. The performance of Pearson correlations improves slightly with increasing $n$, but appears to plateau below 60\%. 

Finally, in case C (ordinal transformation),  polychoric correlations are by far the best option, roughly similar in efficacy to the best performing methods in the ``no transformation'' case. Meanwhile, Pearson, Spearman, and the Gaussian copula struggle in the ordinal case.  These results make sense because polychoric correlation is designed for ordinal variables, whereas Pearson is best suited for Gaussians, and the Spearman and  Gaussian copula approaches likely break down due to the large number of ties. 

Among these simulations, the only case in which PVA with Pearson correlation performs adequately is when the data are truly multivariate Gaussian, where it performs similarly to Gaussian copulas and slightly better than Spearman correlations. This indicates that the standard PVA approaches in the literature are not well suited to non-Gaussian data when we are interested in the latent relationships among variables, and that significantly better performance can be obtained with the alternative methods we propose.

\subsection{Relative explanatory efficiency}
\label{sec:simulations-ree}
A limitation of measuring performance in terms of the proportion of ideal variables selected is that the amount of overlap between a given set of variables and the ideal set of variables is not necessarily indicative of their relative utility, since the correlations among variables may be complex. For example, it could happen that replacing two of the ideal variables with two non-ideal variables would yield a set with greater utility than replacing either one of the two variables alone, even though this would yield a set with a smaller proportion of the ideal variables.

Thus, we also consider the utility of chosen sets of variables in terms of how much of the variance of the omitted variables they explain. Borrowing from the language of ``relative efficiency" of statistical estimators, we introduce the ``relative explanatory efficiency" of $X_S$ versus $X_{S^*}$, defined as
\begin{align}
\label{eq:ree}
    \mathrm{REE}(X_S,X_{S^*}) = \frac{\mathrm{tr}(\mathrm{Cov}(X\mid X_{S^*}))}{\mathrm{tr}(\mathrm{Cov}(X\mid X_S))},
\end{align}
where $X_S$ and $X_{S^*}$ are potentially overlapping sub-vectors of the random vector $X$.
We can interpret $\mathrm{tr}(\mathrm{Cov}(X\mid X_S))$ as the total remaining marginal variance of the variables in $X$ conditional on $X_S$; note that $\mathrm{tr}(\mathrm{Cov}(X\mid X_S)) = \mathrm{tr}(\mathrm{Cov}(X_{S^c} \mid X_S))$ since the variables in $X_S$ have conditional variance 0.
Thus, if $\mathrm{REE}(X_S,X_{S^*}) > 1$ then this indicates that $X_S$ is superior to $X_{S^*}$ in terms of variance explained across all variables, whereas $0 < \mathrm{REE}(X_S,X_{S^*}) < 1$ indicates the reverse.

Our second suite of simulations evaluates the REE of the chosen sets of variables $X_S$ versus the ideal set $X_{S^*}$. We repeat our data-generating set up from \cref{sec:simulations-proportion-selected}, but use REE as the performance metric instead of the percent the ideal variables chosen. Like before, we perform 1000 replicated simulations in which $\bm{\Sigma}$ is sampled from a Wishart distribution, with $p=10$ degrees of freedom and scale matrix $\bm{I}$. Once $\bm\Sigma$ has been converted to a correlation matrix, we produce the latent data $\bm{X}=[X_{ij}]\in\mathbb{R}^{n\times p}$ by sampling $n$ observations from a $\mathrm{MVN}_p(\bm{0}, \bm{\Sigma}$) distribution. The observed data $\bm{Y}=[Y_{ij}]\in\mathbb{R}^{n\times p}$ are obtained by transforming the 5 ideal latent variables, $X_{S^*}$, by the transformations A, B, or C from \cref{sec:simulations-proportion-selected}, where $X_{S^*}$ is determined by performing PVA on $\bm\Sigma$. Finally, the set of ``chosen" variables are determined by applying PVA to $\bm{Y}$ using either the Pearson, Spearman, Gaussian copula, or polychoric correlation matrix. (Polychoric is applied only for transformation C). We compare the chosen set of variables to the ideal set of variables using \cref{eq:ree}. The procedure is repeated for $n\in\{50,150,400,1200,3500,10000\}$.

\subsubsection{Results for relative explanatory efficiency}
 
\begin{figure}[htbp]
\includegraphics[width=\textwidth]{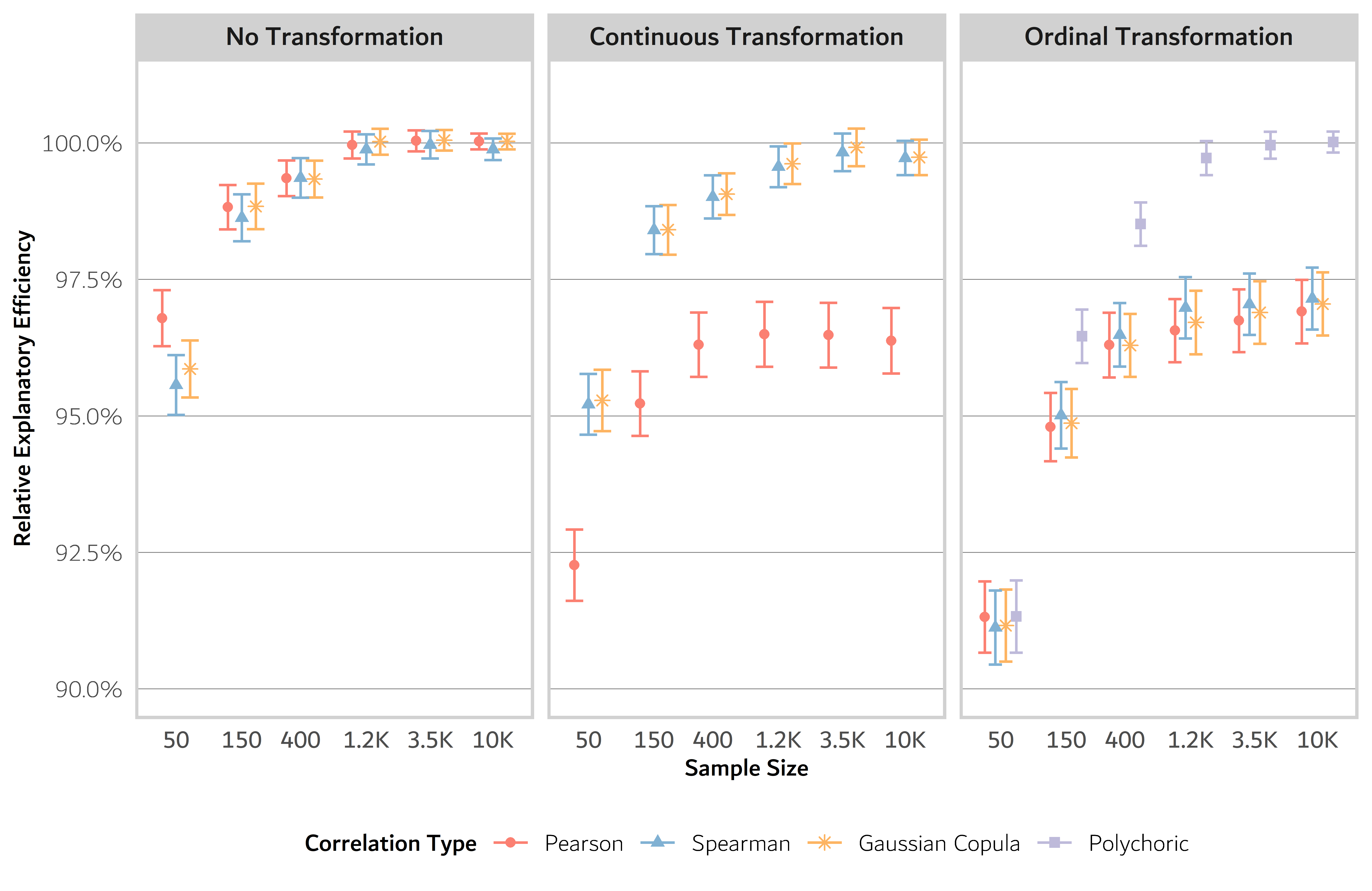}
\caption{\textbf{Relative explanatory efficiency of the chosen variables compared to the ideal variables.} The y-axis is the variance unexplained by the ideal variables divided by the variance unexplained by the selected variables, as a percentage. The value shown is the average over 1,000 simulations, and the error bars show $\pm$ two standard errors. Note that REE $>100\%$ is mathematically possible, since the greedy algorithm is not guaranteed to find the global optimum.}
\label{fig:simulations-ree}
\end{figure}

 Results for these simulations are shown in \cref{fig:simulations-ree}. The y-axis shows the mean REE of the chosen variables relative to the ideal variables, expressed as a percentage. In case A (no transformation), the 
 Pearson correlation method performs slightly better than the Spearman and Gaussian copula method when is $n$ is small. But as $n$ grows large, the mean REE of each of the methods appears to converge to 100\%, and the gaps between the methods disappear. In case B (continuous transformation), the mean REE for the Pearson correlation method plateaus between 95\% and 97.5\%, while the mean REE for the Spearman and Gaussian copula methods again approaches 100\%. In case C (ordinal transformation), the Pearson, Spearman, and Gaussian copula approaches struggle for all choices of $n$, with their mean REEs plateauing below 97.5\%. In contrast, the polychoric method, which is designed specifically for ordinal variables, yields close to 100\% REE with large enough sample size, similar to how the other methods performed when there was no transformation. In general, our findings here are similar to those observed in \cref{fig:simulation-proportion-selected}, where we evaluated performance in terms of the proportion of ideal variables selected rather than REE. However, the performance gaps between methods appear smaller when using REE, since non-ideal sets of variables may still be effective at capturing latent variance.

\subsection{Expanded suite of simulations}
\label{sec:simulations-expanded}
Next, we consider an expanded suite of simulations for generating and analyzing the data. Specifically, we perform simulations under an array of settings in which (1) the number of variables chosen, $q$, can vary; (2) the latent distribution that underlies the data is not necessarily Gaussian, and (3) every latent variable is transformed rather than just the ideal variables.

In this expanded suite of simulations, we again sample a latent correlation matrix $\mathbf{\Sigma}$ that will be used to define an ``ideal" set of variables for that specific $q$. In addition to multivariate Gaussian latent variables, we also consider cases in which the latent variables follow either a multivariate t (MVT) or a multivariate generalized Laplace (MVL) distribution. The MVT and MVL distributions are convenient for this purpose because they are mixtures of the multivariate Gaussian distribution with certain univariate distributions, meaning they have a correlation matrix as one of their parameters. Specifically, if $W\in \mathbb{R}^p \sim \mathrm{MVN}_p(\bm{0}, \bm{\Sigma})$ and $Z\in \mathbb{R}\sim \mathcal{X}^2_{\nu}$ independently, then $W / \sqrt{Z/\nu} \sim \mathrm{MVT}_p(\nu, \bm{0}, \bm{\Sigma})$.  Meanwhile, if $W\in \mathbb{R}^p \sim \mathrm{MVN}_p(\bm{0}, \bm{\Sigma})$ and $Z\in \mathbb{R} \sim \mathrm{Gamma}(r,1)$ independently, then $W\sqrt{Z} \sim \mathrm{MVL}_p(r, \bm{\Sigma})$. These distributions differ from the multivariate Gaussian not only in their marginal distributions but also in their dependence structure, since the mixing with $Z$ introduces tail dependence such that an extreme value in one entry of $\bm{X}$ increases the probability of extreme values in the other entries, which is not the case for multivariate Gaussians. Unfortunately, for these distributions, it is more difficult to compute $\mathrm{Cov}(X\mid X_S)$ and, further, $\mathrm{Cov}(X\mid X_S)$ depends on the specific value taken by $X_S$ rather than just depending on $S$ and $\bm{\Sigma}$. As a workaround, we modify the PVA procedure in \cref{alg:pva} by replacing $\bm{\Sigma}_{-j\mid j}$ with $\bm{\Sigma}_{-j\mid j}^{0} := \mathrm{Cov}(X_{-j}\mid X_j= \mathrm{E}X_j )$ = $\mathrm{Cov}(X_{-j}\mid X_j=0)$, the conditional covariance given that the selected variable takes its mean value. For the $\mathrm{MVT}_p(\nu, \bm{0}, \bm{\Sigma})$ case,
\begin{gather}
\label{eq:mvt}
    \mathrm{Cov}(X_{-j}\mid X_j=0) = \frac{\nu}{\nu - 1}(\bm{\Sigma}_{-j} - \bm{\Sigma}_{-j,j}\bm{\Sigma}_{-j,j}^\texttt{T} / \Sigma_{jj})
\end{gather}
by \citet{Dodge}, 
and for the $\mathrm{MVL}_p(r, \bm{\Sigma})$ case, 
\begin{gather}
\label{eq:mvl}
    \mathrm{Cov}(X_{-j}\mid X_j=0) = (r-0.5)(\bm{\Sigma}_{-j} - \bm{\Sigma}_{-j,j}\bm{\Sigma}_{-j,j}^\texttt{T} / \Sigma_{jj})
\end{gather}
by \citet{Kotz} and \citet{Kozu}. These formulae can be adapted to induce a corresponding version of REE, in which $\mathrm{tr}(\mathrm{Cov}(X\mid X_S=\bm{0}_q))$ replaces $\mathrm{tr}(\mathrm{Cov}(X\mid X_S))$.

The structure of the expanded simulations is similar to previously. Once the correlation matrix $\bm\Sigma\in\mathbb{R}^{p\times p}$ has been sampled ($p=10$), we generate  the latent data $\bm{X}$ by sampling $n$ observations independently from either an MVN($\bm{0}_p$, $\bm\Sigma$), MVT($\nu=2.5$, $\bm\Sigma$), or MVL($r=3.1$, $\bm\Sigma$) distribution. We produce the observed data, $\bm{Y}$, by transforming each of the $p$ variables by the monotonic transformations (A, B, or C) that were defined in \cref{sec:simulations}. Since $p=10$, each of the five transformations is used twice. PVA is then performed to select the best set of $q$ variables, $X_{S}$, using either the Pearson, Spearman, Gaussian copula, or polychoric correlation matrix based on $\bm Y$. We evaluate the performance of each method in terms of $\mathrm{REE}(X_{S},X_{S^*})$, where $X_{S^*}$ is determined by PVA on $\bm\Sigma$. The procedure is repeated 1000 times for each $q\in\{2,3,4,5,6\}$, fixing $n=500$. The results of this procedure are provided below. In addition, for completeness, we repeat our analyses from \cref{sec:simulations-proportion-selected,sec:simulations-ree} in the cases of multivariate t or multivariate Laplace latent data; see \cref{sec:appendix}.

\subsubsection{Results for expanded suite of simulations}

\begin{figure}[htbp]
\includegraphics[width=\textwidth]{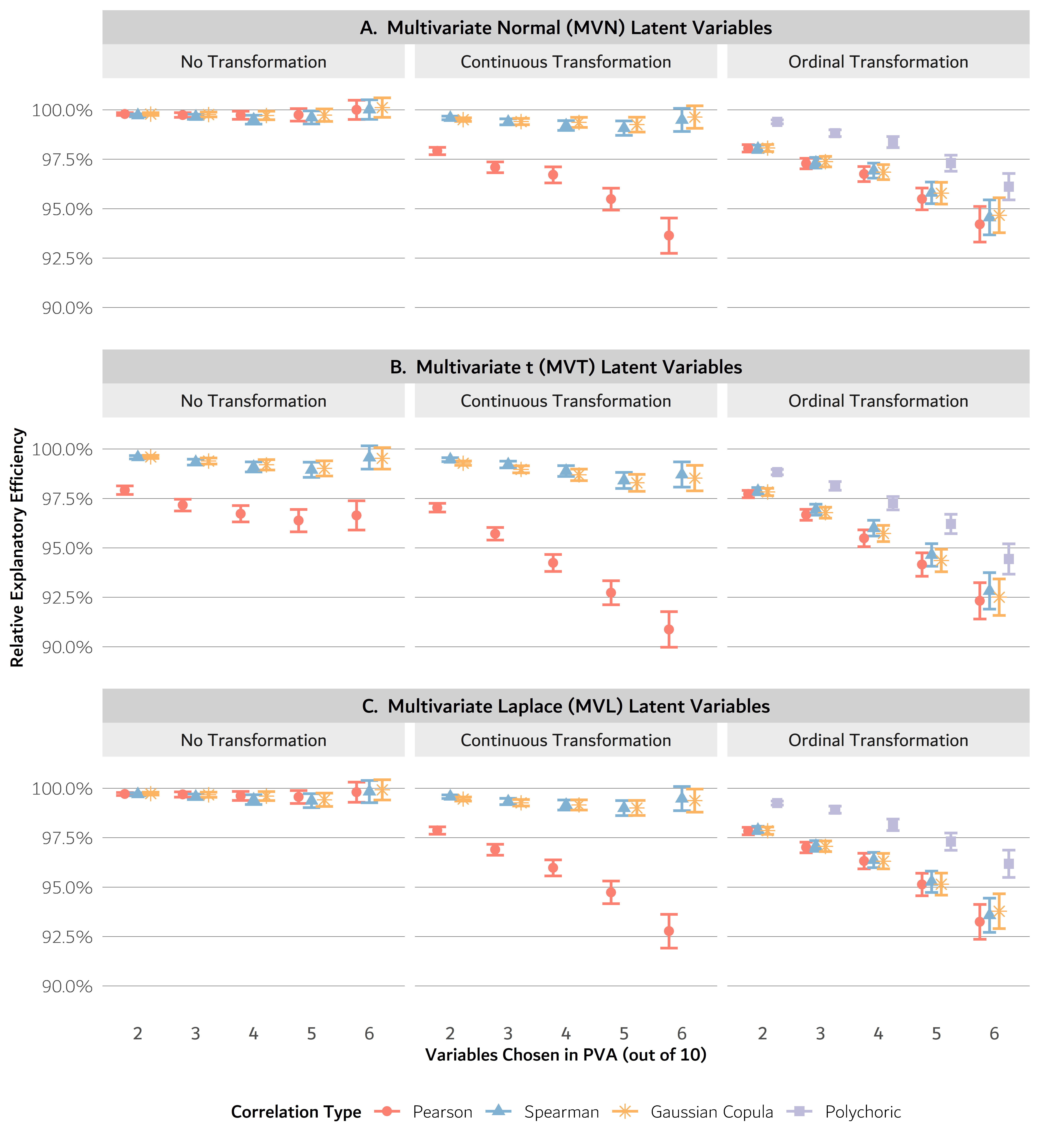}
\caption{\textbf{Relative explanatory efficiency in the expanded suite of simulations.} The y-axis is the variance unexplained by the ideal variables divided by the variance unexplained by the selected variables, as a percentage. The value shown is the average over 1,000 simulations, and the error bars show $\pm$ two standard errors. Note that REE $>100\%$ is mathematically possible.}
\label{fig:simulations-ree-expanded}
\end{figure}

\cref{fig:simulations-ree-expanded} shows the results. Each row of the figure is for a different latent distribution, and each column shows a different transformation type. The y-axis of each plot is the relative explanatory efficiency (REE) of the chosen set of variables compared to the ideal set as determined by running PVA on $\bm{\Sigma}$. We report the average REE over 1000 simulations.

When the latent distribution is multivariate Gaussian (\cref{fig:simulations-ree-expanded}A) and the variables are not transformed (left panel), there are no discernible differences between the three correlation methods (Pearson, Spearman, and Gaussian copula): all three methods perform excellently, with relative efficiencies close to 100\%. However, the performance of the Pearson method is negatively affected by nonlinear transformations, as seen in the ``continuous transformation'' and ``ordinal transformation'' cases (\cref{fig:simulations-ree-expanded}A, middle and right). The Spearman and Gaussian copula methods behave similarly to one another no matter the transformation type, performing well when there are continuous transformations but poorly when the data are ordinal. The ordinal case is best handled by the polychoric method, which clearly outperforms the other methods. Empirically, we observe that REE decreases with the number of variables chosen, indicating that the relative performance of the ideal set compared to the chosen set becomes more prominent as $q$ grows larger.

\cref{fig:simulations-ree-expanded}B and 
\cref{fig:simulations-ree-expanded}C show the results when the latent distribution is multivariate t or multivariate Laplace, respectively. For the multivariate t case, the REEs of the chosen sets of variables compared to the ideal sets of variables are consistently lower than the results in the multivariate Gaussian case, across the board, with the Pearson correlation method appearing to suffer the most. The reduced performance of Pearson compared to the other methods occurred even when the data were not transformed, which emphasizes the lack of robustness of Pearson correlations to violations of multivariate normality. In contrast, the Spearman and Gaussian copula methods were more robust to the change in latent distribution, exhibiting only slightly decreased REEs. The polychoric method was again the best performer when the variables were transformed to be ordinal. Finally, in the multivariate Laplace case, the REEs were only slightly lower across the board compared to the multivariate Gaussian case, with all the methods being affected similarly. In particular, the Pearson REEs were only slightly reduced in the multivariate Laplace case.

\section{Application}
\label{sec:application}


To compare PVA methods on real data, we consider a dataset of longitudinal clinical measurements from individuals with X-linked dystonia parkinsonism (XDP), a rare neurodegenerative disorder occurring in men with maternal ancestry from the island of Panay, Phillipines \citep{LeePasc}. As the name suggests, XDP involves symptoms characteristic of both dystonia (such as muscle contractions, repetitive movements, and abnormal posture) and parkinsonism (such as bodily tremors, slowed movement, and muted facial expressions) \citep{LeeRiv}. XDP is strongly associated with the insertion of a SINE-VNTR-Alu (SVA) retrotransposon in the \textit{TAF1} gene \citep{Aneichyk,Makino}. The SVA retrotransposon contains repeated nucleotide sequences of the form CCCTCT, and greater numbers of repeats of this sequence (that is, larger ``repeat size") are associated with earlier ages of onset \citep{Bragg}. In general, XDP symptoms begin in middle age and then gradually worsen over time. 

In preparation for future clinical trials, \citet{Acuna} conducted a longitudinal study of XDP symptoms,
using Pearson correlation-based PVA to help define a minimal battery of clinical measures that could be used to assess the efficacy of a treatment for slowing or reversing XDP symptoms.
\citet{Acuna} collected data on 29 symptomatic adult males with XDP, visited at 6-month intervals during an 18-month span.  Symptoms were assessed on several published clinical scales, including the Movement Disorders Society Unified Parkinson's Disease Rating Scale (MDS-UPDRS), which measures parkinsonism traits; the Burke-Fahn-Marsden Dystonia Rating Scale (BFM), which measures dystonia-related symptoms; the Eating Assessment Tool (EAT-10), which measures swallowing impairment; and the Communicative Participant Item Bank (CPIB), which measures difficulty communicating. Other measurements targeted the tongue strength and lip strength of the patients, and their ability to make and repeat sounds common to their native language. See \citet{Acuna} for further details.

Our analysis focuses on the 29 XDP-positive individuals who were studied by \citet{Acuna}, each of whom had measurements taken at up to five distinct visits, amounting to 105 visits in total. The variables we consider comprise 102 clinical measures taken at each of those visits; 17 of the variables are continuous and 85 of them are ordinal, with the ordinal variables taking anywhere from 3 to 10 unique levels. The visits are treated independently and serve as the experimental units of the analysis (that is, $n=105$). We handle missingness among the measurements by generating a singly-imputed dataset using predictive mean matching, for which we use the R package MICE \citep{Rubin_1986, mice}. Following the imputation, we run PVA using Pearson, Spearman, Gaussian copula, and polychoric correlations, and compare their performance in terms of finding a small subset of variables that capture as much as possible of the information present in all 102 measures, that is, in terms of defining a minimal battery for future studies. 

\subsection{Application results}

\cref{table:application} shows the first ten variables chosen by performing PVA based on the Pearson, Spearman, Gaussian copula, and polychoric correlation matrices, respectively. We also tried applying polychoric correlations with the continuous variables transformed to be marginally uniform or marginally normal, but this did not make any difference in the first ten variables chosen.

In the first column of \cref{table:application}, we see that PVA based on Pearson correlations chose a variety of variables describing attributes such as speech, swallowing, walking, rigidity, and tremors---covering both dystonia symptoms (the BFM measures) and parkinsonism symptoms (the UPDRS measures). Meanwhile, in the second column, PVA based on Spearman correlations had a strong preference for UPDRS measures over BFM measures; in fact, the only two non-UPDRS measures selected were reported age at onset and EAT-10 Q2. In the third column, we see that the Gaussian copula approach tended to favor UPDRS measures over BFM measures as well, choosing 7 UPDRS measures and only 1 BFM measure. The Gaussian copula was also the only method to select SVA repeat size, which is a known predictor of XDP age at onset. Finally, the fourth column shows results for polychoric correlations, which like the Spearman and Gaussian copula approaches, strongly preferred UPDRS variables instead of BFM variables, although one BFM variable was chosen tenth. Using Gaussian copula or polychoric correlations also resulted in the omission of age at onset, which both the Pearson and Spearman methods did select, albeit close to last. All four methods selected at least one variable describing difficulty eating, though the specific measurement chosen varied.

Since the majority of measures in this dataset are ordinal, we would expect the polychoric method to be better suited at capturing the latent dependency structure of these symptoms.  Therefore, interestingly, the strong preference of the polychoric method for UPDRS over BFM measures suggests that the diversity of XDP symptoms may be better captured by parkinsonism-related metrics rather than dystonia-related metrics, in this particular dataset at least. If forced to choose one of these two measurement scales over the other, this finding suggests that UPDRS might be of greater value for measuring XDP symptoms---an insight that is obscured by using Pearson correlations only.

\begin{table}[htbp]
\small
\centering
\begin{tabular}{lcccc}
  \hline
 Variable & Pearson & Spearman & \makecell{Gaussian\\copula} & Polychoric \\ 
  \hline
BFM-D: Walking & 1 &  &  &  \\ 
UPDRS 3.5: Left hand & 2 & 5 &  &  \\ 
EAT-10, Q1: Swallowing issues led to weight loss & 3 &  & 6 & 4 \\ 
UPDRS 3.18: Constancy of rest tremor & 4 & 3 & 3 &  \\ 
UPDRS 3.3: Neck rigidity & 5 & 6 &  & 6 \\ 
BFM-M: Left Leg & 6 &  &  &  \\ 
BFM-D: Hygiene & 7 &  &  &  \\ 
UPDRS 2.1: Speech & 8 &  & 7 & 9 \\ 
Reported age at onset & 9 & 8 &  &  \\ 
UPDRS-3: Tremor (postural left) & 10 &  &  &  \\ 
UPDRS 2.4: Eating &  & 1 & 1 & 1 \\ 
UPDRS 2.12: Walking and balance &  & 2 & 2 & 2 \\ 
EAT-10, Q2: Swallowing made going out difficult  &  & 4 &  &  \\ 
UPDRS 3.1: Speech &  & 7 &  &  \\ 
UPDRS 3.13: Posture &  & 9 &  &  \\ 
UPDRS 2.7: Handwriting &  & 10 & 10 &  \\ 
UPDRS 3.4: Finger tapping (right) &  &  & 4 & 5 \\ 
UPDRS 3.17: Tremor (left upper extremity) &  &  & 5 &  \\ 
BFM-M: Trunk &  &  & 8 & 10 \\ 
SVA Repeat Size &  &  & 9 &  \\ 
UPDRS 3.17: Tremor (lip) &  &  &  & 3 \\ 
UPDRS 3.17: Tremor (right lower extremity) &  &  &  & 7 \\ 
UPDRS 1.2: Hallucinations &  &  &  & 8 \\ 
\hline
\end{tabular}
\caption{First 10 measures chosen by each PVA method on the XDP data. The numbers 1-10 indicate the order in which the variables were chosen by each method. Blank entries are shown for variables not among the top 10 for each method.}
\label{table:application}
\end{table}

\section{Discussion}
\label{sec:discussion}
The standard approach to principal variables analysis depends on Pearson correlation, which is not well suited for non-Gaussian data.
In diverse simulations, we found that PVA using Pearson correlation performs significantly worse than Spearman, Gaussian copula, and polychoric correlation for capturing the latent dependency structure of non-Gaussian data.
Further, even though the Gaussian copula or polychoric correlation assume a latent multivariate Gaussian distribution,
they still exhibited improved performance over Pearson when the latent distribution was multivariate t or multivariate Laplace, showing that their improved performance is robust to departures from the assumed model.

On the other hand, it is important to bear in mind that this study involved certain assumptions that will not always hold, and in practice, we recommend considering which type of correlation is most relevant to the application at hand.  For instance, our simulation studies assumed that any ordinal variables were discretized approximations to continuous latent variables, and that our primary interest was in the dependence of those latent variables. However, if one is using PVA to select variables that will be included in a linear regression model---where the practical utility of the variables depends on their measurement type---it may be better to choose variables that capture the structure of the observed data rather than the latent data. In a situation like this, Pearson correlation may be preferable to the presented alternatives, since the importance of variables in linear regression modeling depends on their observed linear correlations.

A direction for future work would be to consider other optimality criteria.
Specifically, in this study we considered only the McCabe ``variance-explained'' criterion, in which the objective is to minimize the trace of the conditional covariance matrix of the omitted variables given the selected variables. Other optimization criteria, as well as PCA-based or clustering approaches to PVA, might yield different conclusions about the relative performance of Pearson, Spearman, Gaussian copula, and polychoric correlations for PVA. It would be interesting to study how the observed distribution of the data affects the performance of variable selection in a wider variety of PVA methods and algorithms.


\bibliographystyle{abbrvnatcap}
\bibliography{references}

\appendix
\renewcommand\thefigure{\thesection\arabic{figure}}    
\setcounter{figure}{0}    

\section{Appendix}
\label{sec:appendix}

\begin{figure}[htbp]
\includegraphics[width=\textwidth]{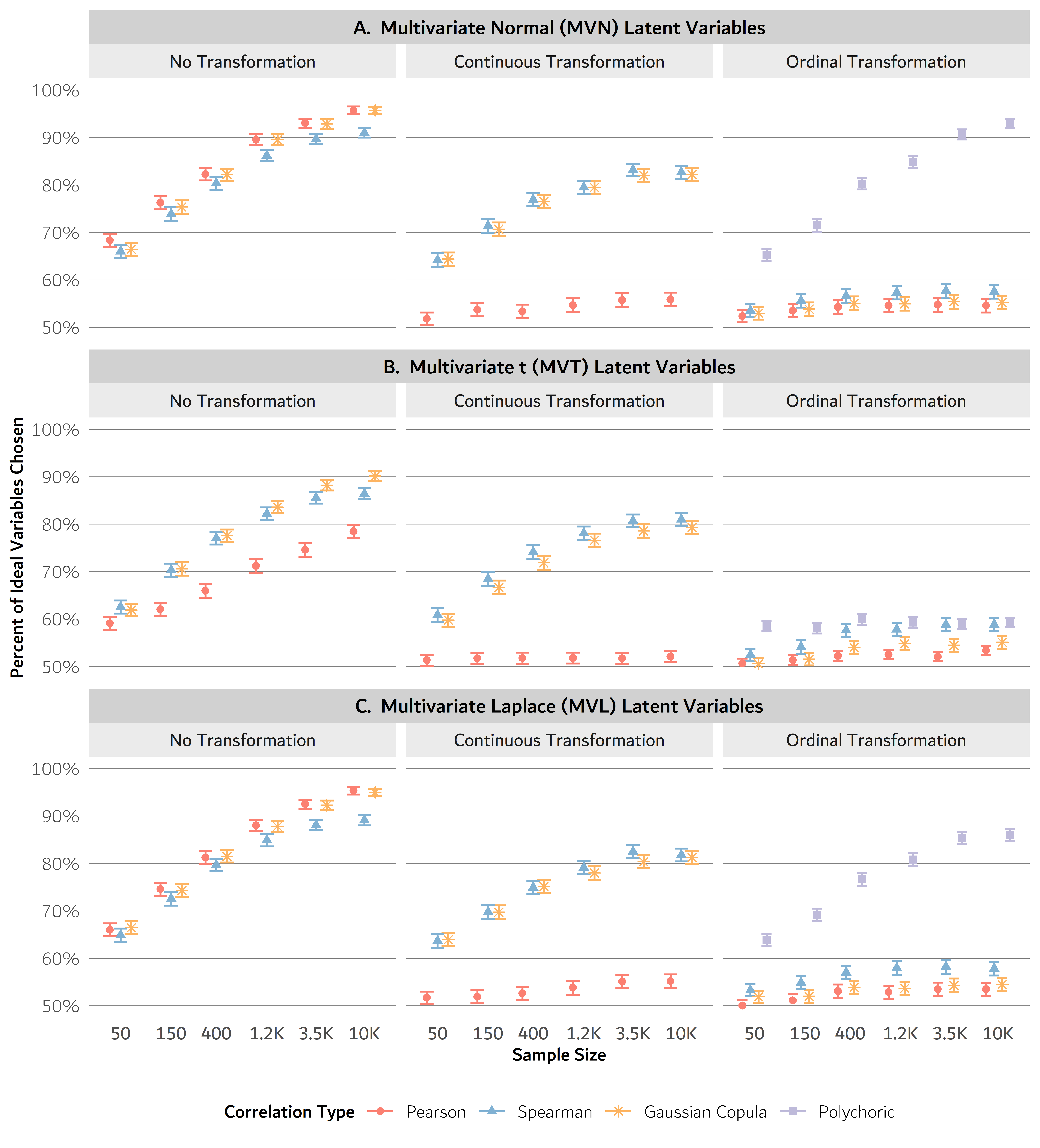}
\caption{\textbf{Proportion of latently ideal variables identified by PVA, varying the latent multivariate distribution.} Value shown is the mean proportion over 1,000 replicate simulations, and error bars show $\pm$ two standard errors. The x-axis is on a logarithmic scale.}
\label{fig:simulations-bestvars-all}
\end{figure} 

\begin{figure}[htbp]
\includegraphics[width=\textwidth]{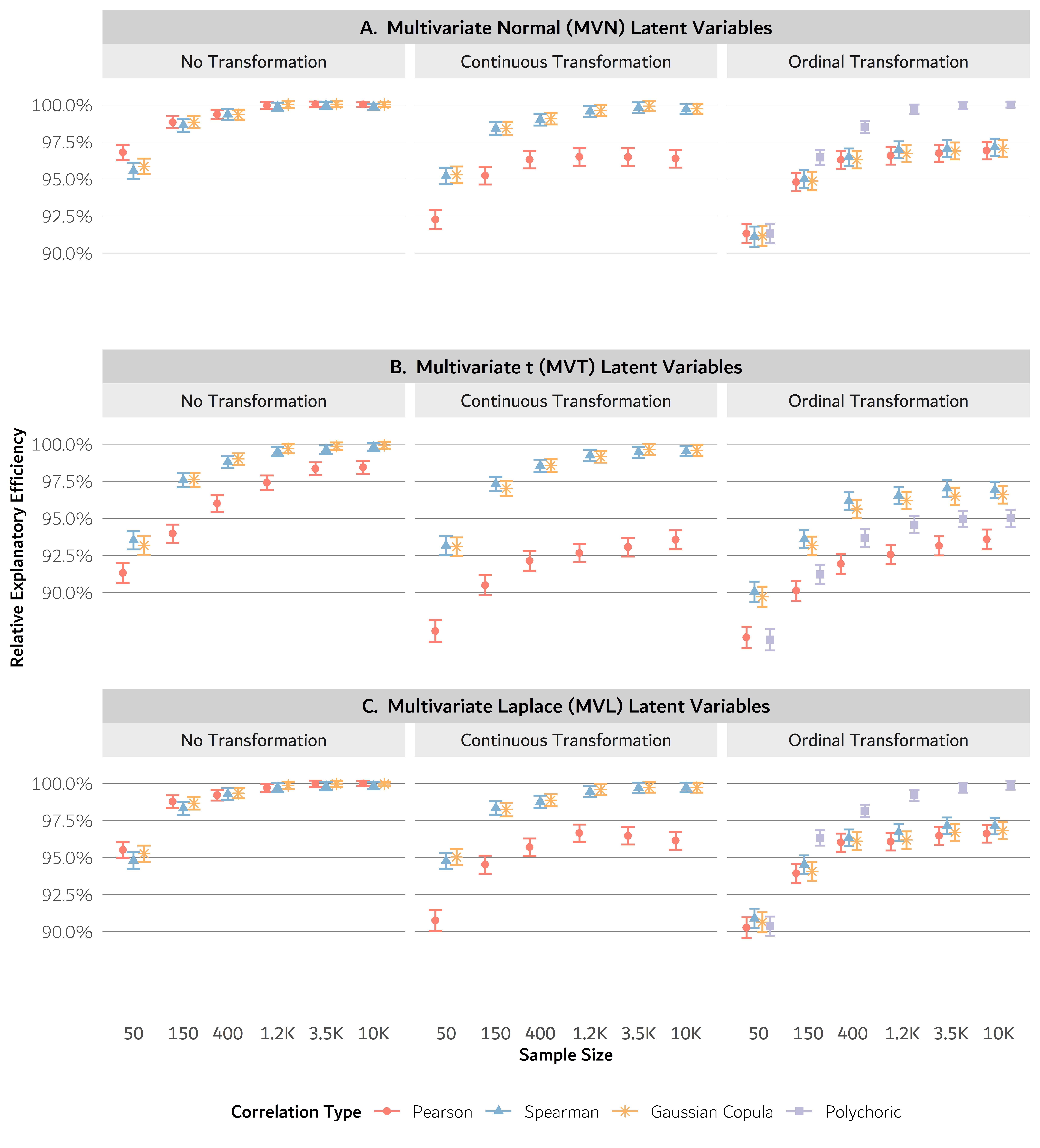}
\caption{\textbf{Relative explanatory efficiency of chosen variables to ideal variables, varying the latent multivariate distribution.} The y-axis is the variance unexplained by the ideal variables divided by the variance unexplained by the selected variables, as a percentage. The value shown is the average over 1,000 simulations, and the error bars show $\pm$ two standard errors. Note that REE $>100\%$ is mathematically possible.}
\label{fig:simulations-ree-all}
\end{figure}

\end{document}